\documentstyle[prl,aps,epsf]{revtex}
\draft
\begin{document}
\twocolumn[\hsize\textwidth\columnwidth\hsize\csname @twocolumnfalse\endcsname

\title{Complex Dynamical Flow Phases and
Pinning in Superconductors with 
Rectangular Pinning Arrays}
\author{C.~Reichhardt and G.T.~Zim\'{a}nyi}
\address{Department of Physics, University of California, Davis, California
95616.}

\author{Niels Gr{\o}nbech-Jensen}
\address{Department of Applied Science, University of California, Davis,
California 95616.\\
NERSC, Lawrence Berkeley National Laboratory, Berkeley, California 94720.}

\date{\today}
\maketitle
\begin{abstract}
We examine vortex pinning and dynamics in thin-film superconductors interacting
with square and 
rectangular pinning arrays for varied vortex densities 
including densities significantly larger than the pinning
density. 
For both square and rectangular pinning arrays, the 
critical depinning force shows maxima at only certain 
integer matching fields 
where the vortices can form highly ordered arrays.  
For rectangular arrays the 
depinning force and commensurability effects 
are anisotropic with a much lower depinning threshold   
for vortex motion in the easy flow directions. 
We find evidence for 
a crossover in pinning behavior in rectangular pinning arrays 
as the field is increased.
We also show analytically,
and confirm with simulations, that for $B = 2B_{\phi}$  
the strongest pinning
can be achieved for rectangular pinning arrangements rather than square 
for one direction of driving force.  
Under an applied driving force  
we find a remarkable variety of distinct 
complex flow phases in
both square and rectangular arrays. These flow phases include
stable sinusoidal 
and intricate pinched 
patterns where vortices from different channels do not mix. 
As a function of the driving force 
certain flow states become unstable and transitions 
between different phases are observed which coincide 
with changes in the net vortex velocities.   
In the rectangular arrays the 
types of flow depend on the direction of drive. 
We also show that two general types of plastic flow occur:  
stable flows, where vortices always flow along the same paths, 
and unstable or  
chaotic flows. 
\end{abstract}

\vskip2pc]
%\vskip2pc
\narrowtext

\section{Introduction} 

Vortex pinning in systems where defects are arranged 
in periodic arrays 
are an ideal system in which to study optimal 
pinning arrangements and commensurability effects 
since the 
properties of the pinning sites, 
such as the size, periodicity of the array, and 
array geometry, can be highly controlled. 
These arrays can be fabricated using nano-lithography 
techniques in which periodic arrangements of 
micro-holes 
\cite{Fiory,Metlushk,Metlushko,Fractional,Wodenweber,Surdeanu,Bezryadin,Moschalkov,Harada,Field,Bael,Welp} 
or magnetic 
dots \cite{Schuller,Jaccard,Fasano,Terentiev,DeLong2,Hoffman,Velez}
of various geometries can be constructed. 
A more recent technique is the use of 
Bitter decoration where the magnetic particles used for the 
initial decoration can act as well ordered arrays of pinning \cite{Fasano}. 
In these systems commensurability 
effects can be  
observed in the form of peaks in the critical current at fields 
where the number of vortices equals an integer or a 
fraction of the number of pinning sites.
At these commensurate fields the  
vortices can form a highly ordered lattice where vortex-vortex 
interactions which reduce the effective pinning are minimized, 
while at 
incommensurate fields the vortex lattice can be disordered and
vortex-vortex interactions make the pinning less effective. 
Direct 
imaging of vortex structures with square pinning arrays 
have been conducted with Lorentz microscopy \cite{Harada} 
and scanning Hall-probes \cite{Field} 
which have  
confirmed that the vortices form highly  
ordered crystals 
at the integer and some fractional matching fields.
These images
along with simulations \cite{Reichhardt} have also shown that
at the different matching fields different types of 
ordered vortex crystals can be
stabilized. In the samples imaged by Harada {\it et al.}, \cite{Harada} it 
was found that beyond 
the first matching field all the pinning sites are occupied
with a single vortex and that
the additional vortices sit in the interstitial regions. The particular 
vortex lattice symmetry
found at the second matching field was square;  
at the third matching field there was an ordered lattice with alternating 
pairs of interstitial vortices; while at the fourth 
matching field the overall lattice
was triangular. The same types of ordered  
vortex lattice crystals were  
also seen in simulations with square pinning arrays and additional
types of stable
vortex crystals were observed for triangular pinning
arrays \cite{Reichhardt}. 
Recent scanning Hall-probe
experiments \cite{Field} have
found various ordered rational fractional fillings
less than the first matching field 
and have also shown that for   
fields greater than the first matching field, vortices can
have both multiple occupancy per pinning site as well as vortices 
that are located in the interstitial regions. 

Most experimental studies of
vortex matter interacting with periodic pinning have 
considered square and 
triangular pinning lattices. Some 
recent studies, however, have been performed with 
Kagome pinning arrays, which produced 
pronounced matching effects at non-integer
matching fields \cite{Jaccard}.  
Periodic arrangements 
with rectangular geometries of magnetic dots 
\cite{Hoffman,Velez} and holes \cite{DeLong2,Koles,WelpR} 
have also been studied recently.  
In 
these systems there can be two periodicities associated with 
the two sides of the rectangular cell, $a$ and $b$. 
In experiments with magnetic dots \cite{Hoffman,Velez}
a field dependent 
crossover in the commensurability effects was observed. 
Sharp low field matching effects occurring at every integer 
matching field were found to cross over to much broader matching effects 
occurring at 
fields where the vortex density 
matches with the periodicity of the short side of the  
rectangular pinning array.
In these systems one would expect an 
anisotropic response, with a lower pinning
or easy flow direction  
through the wide end of the rectangular
cell since the flow
of interstitial vortices will be less impeded
by the vortices pinned by the dots.
Additionally
one would still expect an anisotropic
response due to increased vortex-vortex interactions 
along the short end of the cell at the incommensurate fields which will
lower the pinning force  for fields 
less than the first matching field, as well as for 
fields greater than the first matching field when multiple 
occupancy per pinning site occurs. 
Recent magneto-optical imaging experiments
in samples with rectangular pinning arrays have found evidence 
for anisotropic vortex flow \cite{Koles}. Anisotropic pinning
has also been observed in related systems where the 
individual pinning sites
have an anisotropic geometry \cite{Rectangular}. 

In square pinning arrays 
recent simulations have shown that
a remarkable number of distinct dynamical phases are possible when 
the number of vortices is larger than the number of pinning sites. 
These phases include  one-dimensional flow of interstitial vortices,
random plastic flow,  soliton-like flow along the pinning sites,
and coherently moving elastic flow phases \cite{DrivenShort}.  
Other simulations of vortices in systems with periodic pinning arrays  
have also observed
transitions from different types of  
plastic flow to elastic flow \cite{Marconi,Janin} 
  
Experiments and simulations have also examined the pinning and dynamics 
of vortices through thin channels \cite{Kes,Anders,Besseling}, 
where the vortices moving through the
channels experience a periodic potential created by the ordered lattice
of vortices that are immobile outside of the channels. Peaks in the critical
current are observed where the vortices in the channel can form a 
commensurate ordering. 
Additionally simulations \cite{Besseling} have shown that 
a wide variety of
dynamical phases occur such as soliton like flow, and a periodic 
winding motion of vortices.

Experimentally, evidence for the flow of interstitial vortices in
samples with periodic pinning arrays 
where individual pinning sites are small has been observed 
in transport measurements and current-voltage 
curves \cite{Moschalkov}.  Shapiro steps for the 
1D flow of interstitial vortices between pinning sites with 
an AC and DC applied driving has been
observed in experiments \cite{Bael} and simulations \cite{Shapiro2}.  
Direct imaging of vortices using Lorentz microscopy has demonstrated
the 1D flow of interstitial vortices between pinning sites as well as 
pulse-like motion of vortices along pinning sites and between the pinning sites. 
The pulse or soliton flow of vortices along the 
symmetry directions of the pinning arrays has also been observed with 
magneto-optical imaging \cite{Surdeanu}. 

In systems with random pinning
under an applied driving force the vortex lattice can exhibit different
types of transport behavior such as plastic flow, 
where the vortex lattice is highly disordered or liquid like 
and a portion of the vortices 
can remain immobile 
while other portions 
can tear past. There can also be elastic flow where the moving vortices
keep the same neighbors and the overall lattice 
structure can have a 
crystalline or smectic order. As a function of applied driving force
transitions or crossovers can occur 
between different flow phases. 
Evidence for such transitions has been observed in transport
measurements \cite{Higgins}, voltage noise \cite{Marley}, neutron
scattering \cite{Yaron}, Bitter-decoration \cite{Durate,Pardo}, and 
STM \cite{Troya} and simulations \cite{Koshelev,Moon,Dominguez,Olson}. 
These moving phases have also been  
studied theoretically \cite{Giamarchi}.    
 
In this work we present the results of simulations for vortices in
2D superconductors interacting with square and rectangular pinning arrays.
In particular we examine the pinning characteristics and the dynamical states. 
In previous simulations \cite{Reichhardt} 
only square and triangular pinning arrays were
examined and the dynamical flow states were only examined in 
detail for filling fractions less than $2.25$. In addition 
previous simulations calculated the magnetization curves
but did not
calculate the depinning force directly from current-voltage curves. 
Here, we
examine anisotropic pinning by considering rectangular
pinning arrays for $a/b = 2, 1.6$, and $1$ under applied driving. 
We have also calculated analytically the critical currents in the
$a$ and $b$ direction at the 
second matching fields as a function of $a/b$ and 
find that the critical current
is optimal not for $a/b = 1$ but for a rectangular geometry when 
driven in certain directions. 
For square pinning arrays we find much more pronounced matching effects at
integer and fractional matching fields 
for low applied fields; however, for higher fields the
matching effects are reduced and certain peaks are absent, 
in agreement with recent experiments \cite{Welp}. 
For rectangular arrays with $a/b = 2$ we find that the shape 
and strength of the commensurability effects at the matching fields 
depends on whether the  
driving is in the $a$ or $b$ direction. 
The depinning force 
is much higher along the
long direction. 
We also find evidence 
for a crossover in the commensurability effects at high fields.
Images of the vortices in the
rectangular pinning array show that highly ordered vortex crystals are
stabilized at the matching fields where a peak in the depinning
force is observed, while more disordered crystals are  
formed at the matching fields where peaks are not observed.

Under an applied driving force we show that a remarkable variety of complex   
dynamical phases emerge in square and rectangular arrays. 
These phases depend strongly on the geometry of the 
pinning lattice. 
Despite the complexity of these
dynamic phases 
two general classes of
flow can be identified. The first is 
elastic flow of a mobile sub-lattice of interstitial vortices 
between pinned vortices 
with the vortex motion in 
stable well defined patterns. 
The other type of flow is a chaotic or mixing flow where the 
interstitial vortex
motion is disorderly. The moving channels show a mixing effect, in that the
vortices from one channel move to other channels.
The stable, non-mixing flow phases occur only for
certain integer matching fields and rational machining fields. 
As a function of applied driving force 
dynamical transitions between different types of flow
states are possible which coincide with features in the driving force 
versus velocity curves or voltage-current curves which can be detected 
experimentally.  
 
\section{Simulation} 
We simulate a thin superconductor where the vortices can be
considered as 2D objects interacting with a logarithmic 
potential, $u_{v}=-\ln(r)$, where energy is normalized to
$A_{v} = \Phi^{2}_{0}/8\pi\Lambda$, $\Phi_{0}$ being the flux quantum and 
$\Lambda$ the effective 2D penetration depth for a thin film superconductor.
The normalized overdamped equation of motion for a vortex $i$ is 
\begin{equation}
{\bf f}_{i} = \frac{d {\bf r}_{i}}{dt} = 
{\bf f}_{i}^{vv} + {\bf f}_{i}^{vp} + {\bf f}_{d} = {\bf v}_{i}
\; .
\end{equation} 
The normalized force on vortex $i$ from the other vortices is
${\bf f}_{i}^{vv}=-\sum_{j\neq i}^{N_{v}}\nabla_i U_{v}(r_{ij})$,
where $U_{v}(r_{ij})$ is the effective, re-summed, vortex-vortex
potential between two vortices in a computational cell with
periodic boundary conditions \cite{Jensen}.
Pinning is modeled as attractive parabolic wells with 
\begin{equation}
{\bf f}_{i}^{vp} = -(f_{p}/r_{p})\Theta(r_{p} - 
|{\bf r}_{i} - {\bf r}_{k}^{(p)}|)
{\hat {\bf r}}_{ik}^{(p)}. 
\end{equation}
$\Theta$ is the Heaviside
step function, ${\bf r}_{k}^{(p)}$ is the location of pinning site $k$,
$f_{p}$ is the maximum pinning force, ${\hat {\bf r}}_{ik}^{(p)} = 
({\bf r}_{i} - {\bf r}_{k}^{(p)})/| {\bf r}_{i} - {\bf r}_{k}^{(p)}|$,
and $r_{p}$ is the radius of the pinning sites.
The pinning sites are placed in a rectangular array ($nL_{x},mL_{y}$), where
$n$ and $m$ are integers. The characteristic length, by which all lengths
are normalized, is chosen to be $r_0=\frac{10}{3}r_p$. 
In a typical experiment $L_{x} = 400-900$ nm and the pinning radius is
$125$ nm. The characteristic
time in the normalized equation of motion is $\tau=\eta r_{p}^{2}/A_{v}$, where
$\eta$ is the Bardeen-Stephen friction.
The equation of motion is numerically integrated by an Euler method
using a normalized time step of $dt=0.02$.

The initial vortex positions are obtained from annealing by a 
high temperature, where we add a Langevin noise term 
${\bf f}_{i}^{T}$ to equation (1) such that $\langle{\bf f}_{i}^{T}\rangle=0$
and  $\langle {\bf f}_{i}^{T}(t)\cdot{\bf f}_{j}^{T}(t^{\prime})\rangle=4
T\delta_{ij}\delta(t - t^{\prime})$,
where $T$ is the temperature normalized to $A_v/k_B$, $k_{B}$ being the
Boltzmann constant. We choose an initial temperature
high enough such that the vortices are in a molten state and cool to 
$T = 0$ in 20 increments where we allow 20000 time steps
between each increment. 
After annealing, the driving force is slowly increased 
from $f_{d} = 0$ and the vortex positions and 
velocities are monitored using
$V_{x} = \sum_{i}^{N_{v}}{\hat {\bf x}}\cdot {\bf v}_{i}$ and
$V_{y} = \sum_{i}^{N_{v}}{\hat {\bf y}}\cdot {\bf v}_{i}$. 
In this work we consider simulations where 
the driving is only in the $x$-direction or $y$-direction. 
The depinning force is defined as the force at which
the vortex velocities reach $0.03f_{d}$.

\section{Critical Force at $B = 2B_{\phi}$}
We first consider analytically the depinning threshold in the  
case of 
the vortex lattice driven in the $x$ or $y$ directions at 
$B = 2B_{\phi}$ for arbitrary $L_{x}$ and $L_{y}$. 
At this filling fraction 
the vortex lattice consists of the vortices at the 
pinning sites along with well ordered interstitial vortices located 
between the vortices at the pinning sites. 
We make the assumption that the unpinned vortices form a perfect
rectangular lattice, 
so that they effectively do not interact due to 
symmetry. 
With this assumption we need to 
consider only 
the depinning of a single interstitial vortex moving in the   
periodic potential created by the vortices at the pinning sites. The vortex
will move one-dimensionally along the direction of the drive and experience
the periodic potential created by the vortices at the pinning sites.  
\begin{equation}
\frac{dx_i}{dt} - f_{i}^{vv}(x_i,y_i) = f_{d} \; .
 \end{equation}
The vortex-vortex interaction term $f_{i}^{vv}(x,y)$ can be calculated
using the re-summation method for logarithmically interacting particles in 2D
\cite{Jensen}; the resulting equation of motion for
an interstitial vortex ($i$) moving at $y_i=\frac{1}{2}L_y$ along
the $x$-direction is:    
\begin{equation}
\frac{dx_i}{dt} - 
\frac{\pi}{L_{x}}\sum_{k=-\infty}^{\infty}
\frac{\sin(2\pi\frac{x_i}{L_x})}{ 
\cosh\left(2\pi\frac{L_y}{L_x}(k+ 
\frac{1}{2})\right)-\cos(2\pi\frac{x_i}{L_x})} 
 =  f_{d} \; .
\end{equation}

The maximum depinning force will occur when the summation term is maximum.
In the limit of $L_y\ll L_x$, this happens for $x_i=\frac{3}{4}L_x$,
and the resulting critical force is,
\begin{equation}
f_{x}^{c} = \frac{2\pi}{L_{x}} 
{\rm sech}(\pi\frac{L_{y}}{L_{x}}) \; .
\end{equation}
This expression is accurate for $L_x{\buildrel < \over \sim}1.5L_y$
(the error is about 0.2\% for $L_x=L_y$ and about 5\% for $L_x=2L_y$).
By symmetry, we trivially write the corresponding
critical force in the y-direction, for $L_x\ll L_y$, to be given by,
\begin{equation}
f_{y}^{c} = \frac{2\pi}{L_{y}}{\rm sech}(\pi\frac{L_{x}}{L_{y}})  \; ,
\end{equation}
which is then accurate for $L_y{\buildrel < \over \sim}1.5L_x$.

In figure 1 we plot the critical force in the $x$-direction and $y$-direction
from equations (5) and (6) where $L_{y}$ is kept at a fixed value of $2.0$ and
$L_{x}$ is varied from $4.0$ to $1.25$.
For all increasing $L_{y}/L_{x}$ the critical force in the
x-direction is reduced. This can be understood by considering that the
pinning of the interstitial vortices is caused by the interactions with
the vortices at the pinning sites.
With no driving an interstitial vortex will sit in the center of the
rectangle where the forces from the vortices at the pinning sites
cancel. For decreasing $L_{y}/L_{x}$, the interstitial vortices will have to
move through a 
narrower channel between vortices at the pinning sites in the y-direction  
so the interaction between the interstitial vortices 
and the pinned vortices will increase. 
For driving along the $y$ direction, for 
$L_{y}/L_{x}$ decreasing from 1, the
critical force is reduced. This can be understood by considering that 
as $L_{x}$ is reduced the interstitial vortex interactions
with the pinned vortices in front of the interstitial vortices and those 
behind will increasingly cancel. 
As $L_{y}/L_{x}$ is increased from $1$ the opposite
occurs with the interaction of the vortices that push on the 
interstitial vortices becoming reduced.

Experimentally the behavior of the critical currents as predicted
from equations (5) and (6) can
also be used to determine whether interstitial vortices 
are present at $B/B_{\phi} = 2$ as opposed
to multiple vortices at individual pinning sites. 
In the case of multiple vortices
at the pinning sites the depinning force would be independent of the
ratio of $L_{y}/L_{x}$ since the depinning force  
will not be determined by the 
interactions between the vortices due to
the symmetry of the overall multi-vortex lattice, 
but will instead be determined from the strength of the pinning site only. 
We also show in Fig.~1 the simulation results for the square 
($L_{y}/L_{x} = 1.0$) and rectangular cases ($L_{y}/L_{x} = 2.0$ and
$L_{y}/L_{x} = 0.5$) for driving in the $x$ and $y$ directions showing 
excellent agreement with the predicted values. 

Another experimental signature of interstitial vortices
is the strong difference in the critical force between the
$x$ and $y$ directions at $L_{y} = 2L_{x}$. The critical force 
in the $y$ direction is already more than 
50 times higher than the critical force in the $y$-direction.
Even experiments with a relatively small anisotropy ratio  
such as $L_{y}/L_{x} = 1.25$ should see $f_{c}^{x}/f_{c}^{y} \approx 3$. 
 
These results also show that the maximum pinning can be achieved 
not with square arrays but with rectangular arrays. This enhancement
is only for one of the directions of driving. Figure 1 shows that the
depinning force for $L_{y}=2L_{x}$ 
in the $y$-direction is about $2.5$ times higher than for
the square case. 
In practice the ratio of $L_{y}/L_{x}$ has a finite range in which it can 
be varied due to the finite size of the pinning sites.  

We note that we could calculate analytically 
the depinning force only at $B=2B_{\phi}$, or for any other $B$
values where the interstitial-interstitial vortex interactions effectively
cancel. However,
away from the $B$ values leading to high symmetry interstitial configurations,
the depinning forces should still be anisotropic. 
The depinning force for an interstitial vortex near a 
single interstitial vacancy  
will be  reduced from its value at
$B=2B_{\phi}$ by a quantity  $\propto 1/L_{x}$ if driven in the
$x$ direction and $\propto 1/L_{y}$ if driven in the $y$ direction since 
the vortex-vortex force goes like $\propto 1/r_{ij}$. Similar arguments
for an anisotropic depinning force can
be made for additional interstitial vortices added to the $B=2B_{\phi}$ 
vortex configurations. 

\section{Dependence of Depinning Force on $B/B_{\phi}$ in a Square Array}

In Fig.~2 we show the dependence of the critical depinning force  
as a function of $B/B_{\phi}$ for a system with a square pinning array. Here
sharp  
peaks in $f_{p}^{c}$ can be seen 
at $B/B_{\phi} = 1, 2, 3, 4$, and $5$.
Smaller peaks are seen for $B/B_{\phi} = 6$ and $8$. At 
$B/B_{\phi} = 7$ there is no evidence for a peak. Another 
interesting feature is that for 
$6 < B/B_{\phi} < 8$ the critical force remains at an intermediate
value which is higher than the lowest critical current values for 
$B < 5B_{\phi}$. This is a similar trend to that observed by
Metlushko {\it et al.} \cite{Welp} 
who claim that interstitial vortices are present for $B/B_{\phi} > 1$ 
due to the small size of the pinning sites. 
The peaks observed in Fig.~2  
are much stronger then those observed 
in magnetization measurements in simulations
with flux-gradient driven vortices with short range 
interactions \cite{Reichhardt}.  
In that work 
commensuration enhancements were not seen at  
$B/B_{\phi} = 3, 6$, or $7$. The  vortex configurations  
observed 
at the matching fields are the same as those for Ref [26].    
Another feature in Fig.~2
is that the 
height of the peaks for $B < 5B_{\phi}$ shows variations with
the largest peak at $B = 4B_{\phi}$ when the vortices form
a triangular lattice as seen in simulations and experiments.  
Some experiments with square
and magnetic dots have 
observed strong matching effects  at 
{\it every} matching field \cite{Schuller} suggesting the
presence of multi-vortex states in these systems.  The inset of 
Fig.~2 shows the depinning curve for a system with the same parameters
as Fig.~2, but with the pinning sites in a random arrangement.
Here the depinning force decreases with increasing field and there
are no peaks at the matching fields.   

In Fig.~2 
commensurability peaks at 
the fractional fields $1/2, 3/2$, and $5/2$ can also be observed. Peaks in
the critical current at fractional
$B/B_{\phi}$ have been seen in experiments 
with particularly pronounced peaks at $1/2$ and $3/2$ while weaker 
peaks were seen at $1/4, 1/5$, and $1/16$ \cite{Metlushko}.  
The vortex configurations and dynamics at fractional matching fields 
are studied in detail elsewhere. 
We do not observe any particular fractional matching    
except  at $B/B_{\phi} = 8.5$ where a clear peak in the critical
current is observed. 
Such a commensuration peak has not been seen in previous simulations or 
experiments.
In section V (c) we show that a stable non-mixing flow state 
occurs at this field.  

\section{Dynamical Flow States for Square Pinning Arrays}
\subsection{1D Flow States at $B/B_{\phi} = 2, 4$ and  $9$} 
Figure 3(a) shows the flow states 
just above depinning for $B/B_{\phi} = 2$ where
the motion consists of the straight 1D flow of interstitial vortices between 
the pinned vortices. This type of flow was also seen in simulations at the
same field for vortices with bulk interactions. 
In Fig.~3(b) for $B/B_{\phi} = 4$ 
a similar 1D straight vortex flow is observed. Here  
the vortices at 
the pinning sites also remain pinned and 
there are additional interstitial vortices located between the 
pinning sites in the $x$ direction that also remain pinned. The mobile 
interstitial vortices can move 
in a straight unhindered 1D path while the immobile interstitial vortices
cannot
move in 1D paths without entering a pinning site or coming close to the
vortices located in the pinning sites. 
In Fig.~3(c) at 
$B/B_{\phi} = 9$ the same type of 1D 
interstitial flow as in Fig.~3(a) and Fig.~3(b) is observed 
but in this case there are two
mobile rows of vortices between the pinning sites and 
two immobile 
interstitial vortices between vortices at the pinning sites. In
Fig.~3(a,b,c) the vortex motion can be seen to be elastic with respect  
to the mobile vortices in which the moving vortices keep their same 
moving neighbors. 
Further, the vortices always flow in the same paths. 
At fields with a fraction of $0.05$ higher or lower than the   
matching fields in Figs.~3(a,b,c) 
the initial vortex motion occurs
at the location of the extra vortex or vacancy 
in the ordered interstitial lattice. The 
depinning occurs at a lower driving force than
that at which the commensurate vortex configuration depins.
The 
flow of these extra interstitials or vacancies will again be 
in a 1D path along the direction of drive; however, the motion
is not continuous but occurs in a soliton or pulse fashion with 
the extra interstitial or vacancy exchanging places with
pinned vortices as it 
propagates. A similar soliton like motion of vortices along the pinning
sites has also been observed previously  
in simulations \cite{Reichhardt} and experiments \cite{Surdeanu}.

\subsection{Sinusoidal Flow States At $B/B_{\phi} = 5, 8$, and $8.5$} 
In Fig.~4(a) we show the flow states at $B/B_{\phi} = 5$ which shows
sinusoidal flow states of the interstitial vortices. Here the
vortex motion is not strictly 1D in the direction of drive but shows
a periodic motion in the transverse direction as well. In addition 
the vortices flow in the same paths and vortices from one 
channel do not mix with vortices in the other channels. 
In Fig.~4(b) a similar sinusoidal flow as seen at $B/B_{\phi} = 5$
is observed with an additional
square sub-lattice of pinned interstitial vortices
at the center of the pinning plaquette similar to the vortex configuration
at $B/B_{\phi} = 2$. For higher drives the immobile 
interstitial vortices depin and the
vortex lattice enters a new flow phase. 
In Fig.~4(c) for $B/B_{\phi} = 8.5$, where
a small peak in the critical current is seen, an interesting flow state is 
observed where there is a pinned interstitial lattice for every other 
interstitial vortex in the middle of the pinning array plaquette. Additionally
there are two immobile vortices between pinning sites in the 
$x$-direction. The vortex flow is a combination of 1D flow and sinusoidal 
flow with the sinusoidal flow occurring at every other plaquette 
containing a pinned interstitial vortex.  

\subsection{Bistable Flow States in the Vicinity of $B/B_{\phi} = 3$}
Fig.~5(a,b) shows typical vortex flow patterns observed for fields 
$2.75 < B/B_{\phi} < 3.5$. Here the 
interstitial vortices 
move in periodic meandering paths around the pinned vortices.   
Interestingly the direction 
of the vortex flow is {\it not} in the direction of the driving force but at 
45 degrees in either transverse direction. 
Since either direction is equivalent  
the flow jumps between the two states as seen in the 
measured transverse $V_{y}$ velocities.
In Fig.~6,  $V_{y}$ shows a small amplitude periodic component caused by the
winding nature of the vortex flow, along with low frequency 
large amplitude jumps of $V_{y}$ from positive 
to negative indicating that the net vortex
motion direction is changing. 
For different system sizes the characteristic of the flow paths
remains the same. 
Experimentally this flow state can be observed with transverse noise
measurements. 
For stronger drives there is a transition to a more disordered flow in the
direction of the driving force.

\subsection{Disordered Vortex Flow States and Noise} 
In Fig.~7(a,b,c) we show the vortex flow states at various incommensurate
fields showing varying degrees of disordered or mixing flow states.
In 
Fig.~7(b) the vortex trajectories for $B/B_{\phi} = 4.95$
are plotted showing that the
general features of the stable channel flow from $B/B_{\phi} = 5$  
are still present. There are, 
however, now some vortices that can be seen to
jump between adjacent channels. 
In Fig.~7(a) for $B/B_{\phi} = 4.5$ the vortex 
trajectories
are more disordered and no discernible preferred channels are
visible. This same type of flow is seen for $B/B_{\phi} = 5.5$. 
There are regions near the pinning sites where repulsion from the
vortices located in the pinning sites keeps other vortices from approaching 
closer than a certain distance. 
In Fig.~7(c) 
we show the vortex trajectories for 
$B/B_{\phi} = 7$ showing that 
the disordered flow is again similar to what was seen
at $B/B_{\phi} = 4.5$. We note that this is the matching field at which
a peak was absent in the depinning force.
In general we observe that the non-mixing flows
occur only near certain integer or fractional matching fields where the 
initial pinned vortex lattice has an ordered state. The flows then 
become increasingly disordered as the fields move away from these 
matching configurations. 

To further compare the 
experimentally measurable noise signals of the
ordered and disordered vortex states we plot in Fig.~8 the time
series and the corresponding Fourier transforms (FT's) 
of the average vortex velocity $V_{x}$ for a fixed applied
driving force. Each data set is analyzed for 30,000 MD steps. 
In the figures we only show a portion of the time series for clarity.
For $B/B_{\phi} = 4$ (Fig.~8(a,b)), where the flow was seen to occur in 
1D channels as seen in Fig.~3(b), 
only a single periodic component 
and higher harmonics can be seen. For 
$B/B_{\phi} = 4.5$ (Fig.~8(c,d)) 
where the flow was disordered as seen in 
Fig.~7(a), a more random 
noise signal is observed and the FT shows a broad spectra.    
At $B/B_{\phi} = 5$ (Fig.~8(e,f))
where an ordered flow state was observed 
as seen in Fig.~4(a), a  
clear single frequency signal is again observed.
In Fig.~9 we show the time series and FT for $B/B_{\phi} = 3$ where
more complicated vortex trajectories were observed as seen in
Fig.~5(a,b). Here the  signal shows several different periodicities. 
In general we find
periodic signals where orderly or partially ordered flows are
observed, and broad noise spectra 
where incommensurate or disorderly flows occur.

\subsection{Longitudinal and Transverse Noise}

In Fig.~10(a,b) we show the time series for both the sum of the
velocities in the $x$-direction and $y$-direction simultaneously for
(a) $B/B_{\phi} = 4$ and (b) $B/B_{\phi} = 5$. For
$B/B_{\phi} = 4$ longitudinal velocities show the periodic signal as 
shown earlier while the transverse velocities are zero as the vortices are
moving strictly in the $x$-direction in 1D channels as seen in 
Fig.~3(b). For $B/B_{\phi} = 5$ both the longitudinal and  
transverse velocities show a pronounced periodic signal as the vortex
trajectories show a sinusoidal flow as seen in Fig.~4(a). These 
results suggest that the features of the flow, such as whether it is
winding in the transverse direction or strictly 1D, can be probed
experimentally with transverse noise measurements.  
The periodic time signals at the matching fields can also be probed with
an applied AC drive superimposed on a DC drive. Shapiro steps can be
observed, as recently determined in experiments and simulations at 
$B/B_{\phi} = 2$. Where the vortex flows are highly ordered one would
then expect Shapiro steps \cite{Shapiro2}, while at the disordered flow
states the phase-locked steps would be absent or strongly reduced.  

\section{Dependence of Depinning Force on Field For Rectangular Pinning
Arrays}

\subsection{Rectangular Pinning Array with $L_{x}/L_{y} = 2$}

In Fig.~11(a,b) we show the dependence of the critical depinning force on 
$B/B_{\phi}$ for systems with similar parameters as studied for the
square array
with $L_{x} = 2L_{y}$ for driving in the $x$-direction and
$y$-direction. Here $f^{c}_{x}$ is much higher than $f^{c}_{y}$ for
all $B>B_{\phi}$. For $ B/B_{\phi} < 6$ 
commensurability peaks are observed for driving in both
the $x$ and $y$ directions; 
however, the peaks in the $x$-direction do not 
show the sharpness of the peaks 
seen for the $y$-direction. The peaks for both driving directions 
are much broader for $B/B_{\phi} > 5$. There are no clear peaks at 
$B/B_{\phi} = 7$ or $9$; however, there is some evidence for a 
drop-off in the depinning force right after these fields. 

\subsection{Vortex Configurations for $L_{x}/L_{y} = 2$}

In Fig.~12(a-i) 
we show the vortex configurations for the various matching fields. 
The vortex configurations at $B/B_{\phi} = 1$ and $2$ (Figs.~12(a,b))  
are rectangular in order. For $B/B_{\phi} = 3$ 
(Fig.~12(c)) the vortex 
crystal has a diagonal ordering.
At $B/B_{\phi} = 4$ (Fig.~12(d)) a square vortex lattice is stabilized. 
At $B/B_{\phi} = 5$ (Fig.~12(e)) the vortex lattice does not form a 
simple rectangular lattice but an ordering is still present as indicated
by the unit cell. We note that at $B/B_{\phi} = 5$ a peak or anomaly in 
the depinning force is not observed. For $B/B_{\phi} = 6$ and $8$, 
ordered distorted hexagonal vortex lattices are stabilized. At
$B/B_{\phi} = 7$ and $9$ the vortex lattice is disordered
which again corresponds to there being no peaks in 
the critical depinning force.
These
results indicate that at the matching fields where an ordered lattice with
a simple rectangular or hexagonal ordering can occur, an enhancement
in the critical current can be observed.

\section{Dynamical Flow States for Rectangular Pinning Arrays}

In Fig.~13(a-f) we show the flow states at depinning in the
rectangular arrays for $B/B_{\phi} = 2, 3, 4, 5, 6$, and $9$,
respectively, for driving along the long side or $x$-direction. 
1D flows that are similar to those seen in the square arrays occur for
$B/B_{\phi} = 2$ and $6$ (Fig.~13(a) and Fig.~13(e)). Disordered flow phases
occur for $B/B_{\phi} = 3$, $4$ and $9$
(Fig.~13(b), Fig.~13(c) and Fig.~13(f)) where only minimal anomalies were
seen in the depinning force. A disordered phase is also observed for 
$B/B_{\phi} = 7$ (not shown) where there was
again no anomaly in the depinning force. 
An ordered flow phase is observed for $B/B_{\phi} = 8$ 
which is described in the next section where
a peak in the depinning is observed. 
At $B/B_{\phi} = 5$ (Fig.~13(d)), where 
a static ordered lattice was stabilized but an anomaly in the depinning
force was not observed, we find an interesting 
periodic winding flow of interstitial vortices. Also pairs of immobile 
interstitial vortices 
are located between every other pair of pinning sites. 

\subsection{Pinched Sinusoidal Flow For $B/B_{\phi} > 5$ for 
Driving Along the Short Edge}     
A new
type of stable flow phase that appears for the rectangular arrays 
for $B/B_{\phi}> 4$ at higher drives is a
{\it pinched sinusoidal} flow phase such as seen in 
Fig.~14(d) for $B = 8$. We follow the development of this phase
in Fig.~14 for $B/B_{\phi} = 8$ where the initial depinning occurs
through the almost 1D flow of interstitial vortices where two interstitial 
vortices remain pinned between the pinning sites as 
seen in Fig.~14(a) where $f_{d} = 0.1$.
There is additional
slower periodic wandering of the moving vortices in the $y$-direction
that broadens the channel. This periodic transverse wandering of the
channels can also be seen to push the immobile interstitial vortices
in the transverse direction. 
The trajectories from the longitudinally moving
vortices are pinched where the 
moving vortices
are directly adjacent to the pinning sites. 
In Fig.~14(b) with $f_{d} = 0.2$ we show that for increased driving 
forces there is a transition to a disordered state as the immobile
interstitial vortices begin to depin. For $f_{d} = 0.3$ 
(Fig.~14(c)) the flow starts 
to reorganize to the pinched sinusoidal flow which becomes fully
developed for $f_{d} = 0.325$ as seen in Fig.~14(d).  

In the pinched sinusoidal flow the vortices 
show a stable channel flow similar to the 
sinusoidal flows seen in the square arrays; however there
is a pinched feature as the vortices flow adjacent to the
pinning sites. 
As the vortices move through the pinched 
area only one vortex goes through the intersections
at a time with a vortex from the upper and lower part of the branches 
alternating. For the
other matching fields for $B/B_{\phi} > 4$
the same type of flow patterns are observed. 
The same flow patterns are 
observed at fields not too far from commensuration with the mobile rows 
carrying a varying number of vortices. 

\subsection{Additional Dynamical Flow Phases for $B/B_{\phi} < 4$} 
In Fig.~15(a) we show an interesting type
of dynamically induced rotational motion
ordered flow phase that occurs 
at and near $B/B_{\phi} = 2.5$ for driving along the long side of the
pinning lattice. 
The depinning is
due to the 1D flow of a portion of the interstitial vortices 
which move between   
every other pinning row. 
Another portion of interstitial vortices cannot pass through the 
pinned vortices and show a remarkable 
paired collective rotational motion.
The rotations occurs as a pair of interstitial vortices rotate in opposite 
directions as the 1D moving interstitial vortices pass. 
Additionally every other rotating pair is out of phase with the adjacent
rotating pairs. The rotational motion occurs due to the fact that the 
rotating interstitial vortices are located 
in a shallow potential well created
by the vortices located at the pinning sites. As the 
 ordered 1D moving vortices move past they push the 
interstitials vortices that are not moving longitudinally.
Although these pairs do not show a net motion in the
$x$-direction they are still taking part in the dissipation and slow down
the 1D moving vortices. 
For higher drives the rotating interstitials can be dislodged from 
their orbits and the flow becomes more disordered. 

In Fig.~15(b) we show the ordered flow phase that occurs for 
$B/B_{\phi} = 4.5$ where the initial depinning occurs through
the sinusoidal flow of a portion of the interstitial vortices. Unlike 
the previous sinusoidal flows observed in this case the channels 
weave across the pinning sites rather then strictly between two
rows of pinning sites. For higher drives the other immobile vortices
depin and the flow enters a disordered phase. 

\subsection{Flow Phases For Driving Along Long Edge} 
In Fig.~16(a-f) 
we show the flow states just above depinning for 
driving along the short side of the rectangular pinning array for
$B/B_{\phi} = 2, 3, 4, 7, 8$, and $9$, respectively.
The motion at $B=2B_{\phi}$ (Fig.~16(a)) 
is 1D where a single channel of 
interstitials is moving while for $B/B_{\phi} = 8$
(Fig.~16(e)) three rows of vortices
are flowing and an immobile interstitial vortex is located between the
pinning sites. An almost 1D flow is seen at $B/B_{\phi} = 7$
(Fig.~16(d)) in 
which 3 rows of interstitial vortices are moving. We note that 
for $B/B_{\phi} = 7$ a peak or matching anomaly was not present. 
The vortex configurations for $B/B_{\phi} = 7$ shows  
that along the immobile row of vortices there are
16 vortices with 8 located in the pinning sites while the mobile rows
each contain 13 vortices so that the moving vortices are not commensurate
with the periodicity 
of the potential created by the immobile vortices. 
For the $B/B_{\phi} = 8$ matching field where a peak was seen, the
vortex configurations again show 16 vortices along the immobile row but
the moving rows each contain 16 vortices which can be commensurate with
the immobile vortices. The flow paths for 
$B/B_{\phi} = 3$ (Fig.~16(b)) and 
$B/B_{\phi} = 4$ (Fig.~16(c)) 
each show two rows of moving interstitial vortices
in a sinusoidal flow with the flow at $B/B_{\phi} = 4$ showing a 
a larger amplitude. For $B/B_{\phi} = 9$ (Fig.~16(f)) where there was an
absence of the peak in the depinning force the flow is mostly disordered
with some remnant of the flow phase seen at $B/B_{\phi} = 8$.  

 In Fig.~17(a,b) we show the flow phase at $B/B_{\phi} = 5$, and 
$6$, respectively,  for
driving along the short edge. The flow at $B/B_{\phi} = 7$ shows a
remarkable {\it braiding} flow where the moving interstitial vortices 
flow in a crossing pattern. A moving vortex that starts in the region
almost underneath the pinning sites will move at 
an angle until it reaches
the opposite side of the channel two pinning sites up while other vortices
cross in the opposite direction. At $B/B_{\phi} = 6$ a similar flow to
that seen at $B/B_{\phi} = 5$ is observed; however, the motion is not
perfectly ordered.

\section{Conclusion}
We have analyzed numerically the vortex pinning and dynamics in square and
rectangular pinning arrays for thin film superconductors. We predict
an anisotropic critical current for the rectangular pinning arrays and 
calculate analytically the ratio of the critical depinning forces for 
different aspect ratios at $B/B_{\phi} = 2$. These results show that
as the critical current is enhanced in one direction it is reduced in the 
other direction. The maximum critical current can be achieved for 
driving in certain directions of the rectangular pinning array. Numerical
simulations also find that the anisotropy in the critical depinning
force occurs for $B/B_{\phi} < 4$ and is less pronounced for higher 
fields. 
In the case of the square
pinning arrays we observe pronounced commensurability effects at 
most integer matching fields 
and a missing matching peak at 
$B/B_{\phi} = 7$ in good agreement with recent experiments.  

In the rectangular pinning arrays with an aspect ratio of 2:1 we observe 
an anisotropic depinning threshold in which the easy flow direction shows 
an overall lower depinning force but sharper and more pronounced commensuration
effects while in the other direction the 
overall depinning force at both the commensurate and 
non-commensurate fields is higher, but the
matching effects are considerably reduced. We observe 
integer matching effects in both driving directions 
up to $B/B_{\phi} = 5$, after which only every other matching field shows
an enhanced critical current. The vortex configurations show that
at the matching fields where enhanced critical currents are observed 
an ordered vortex arrangement is formed while at the matching fields
where there is no enhancement the vortex arrangement is disordered.    

For both the square and rectangular arrays we find that
a remarkable variety of 
intricate dynamical flow phases can be realized and in 
general two classes of flow phases can be observed: stable channel
flow, where vortices flow in the same paths in identifiable channels and 
vortices from one channel do not mix with vortices in another 
channel; and
a disordered or mixing flow, where vortices from different channels mix or
no identifiable channels occur. The particular flow states 
we observe include a 
1D flow of interstitial vortices. 
For the square arrays at 
$B/B_{\phi} = 5, 6$, $8$ and $8.5$ a sinusoidal flow occurs where 
a portion of the interstitial vortices are moving in winding paths while 
another portion of interstitial 
 vortices remain pinned along with the vortices
at the pinning sites. For increasing 
drives the immobile interstitial vortices
can depin and the vortices can enter a more disordered flow phase.  
Near $B/B_{\phi} = 3$ a remarkable bistable flow phase is observed where the
vortex flow is not in  the direction of drive, but is alternating 
from the $+45$ degree direction to the $-45$ degree direction. This motion 
can also be seen in the transverse vortex velocities.  
Away from commensurability these ordered flow
phases become increasingly disordered. We also show that 
these phases can be probed experimentally with noise
spectra. In ordered 
phases the noise spectra shows distinctive narrow-band features, while the
disordered flows show a broad spectra. 
In addition we show that with transverse noise measurements
the 1D and sinusoidal flows can be distinguished where the sinusoidal
flows will show a narrow band transverse signal.      

In the rectangular arrays several of the phases observed in the square
arrays are also observed. A new
ordered flow phase that occurs for driving in the short direction 
for a wide range of fields for high driving is a braided channel flow. 
At the matching fields the braided flow is uniform 
with each channel carrying the same number of vortices while at 
incommensurate fields the channels can carry different numbers of 
vortices. 
The signature of the onset of this braided flow can be seen in the
current-voltage characteristics.  

\begin{center}
{\bf Acknowledgments}
\end{center}
We thank C.J.~Olson for critical reading of this manuscript, and
I.K.~Schuller, J.~Martin, L.E.~DeLong, L.~Van Look, 
V.~Metlushko, V.V.~Moshchalkov, 
A.~Hoffman, F.~Nori, and R.T.~Scalettar for 
useful discussions. This work was supported by the Director, Office of
Advanced Scientific Computing Research, Division of Mathematical, Information
and Computational Sciences of the U.S.~Department of Energy under
contract number DE-AC03-76SF00098 as well as CLC and CULAR (Los Alamos
National Laboratory).

\begin{figure}
\center{
\epsfxsize=3.5in
\epsfbox{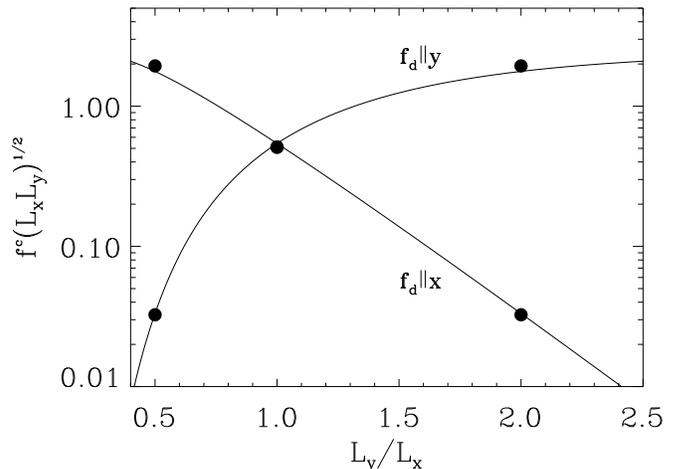}}
\caption{
Critical depinning force for driving in the $x$-direction 
and $y$-direction
vs the aspect ratio of the rectangular pinning array for $B=2B_{\phi}$
as predicted by Eq.~(5) and Eq.~(6)
for a constant $L_{y} = 2.0$ with $L_{x}$ varied from $4.0$ to $1.25$.
The filled circles are the results from simulations.} 
\label{fig1}
\end{figure}

\begin{figure}
\center{
\epsfxsize=3.5in
\epsfbox{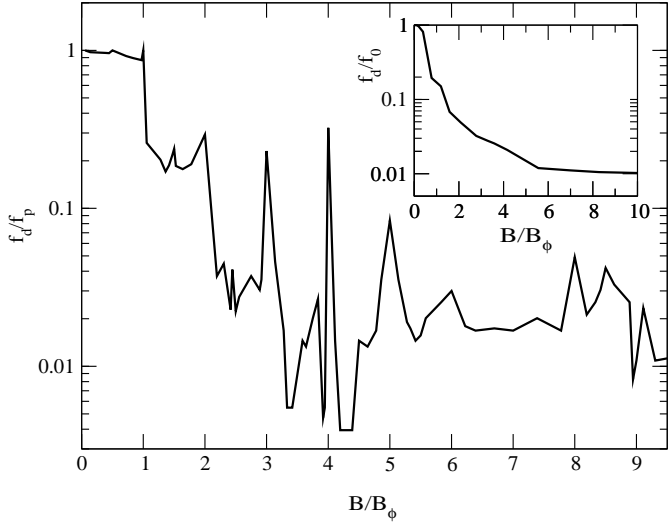}}
\caption{
Critical depinning force vs the field $B/B_{\phi}$ in a system with
a square pinning array for $nL_{x} = nL{y} = 12.0$.
Peaks in the depinning force can be seen at 
most of the matching fields with a clear missing peak at 
$B/B_{\phi} = 7$. In addition some clear peaks can be seen at the fractional
matching fields $B/B_{\phi} = 1/2$, $3/2$ and $5/2$. The inset shows the
depinning force vs $B/B_{\phi}$ for a system with the same parameters
but with the 
pinning sites in a random arrangement showing the
absence of matching effects.}
\label{fig2}
\end{figure} 

\begin{figure}
\center{
\epsfxsize=3.5in
\epsfbox{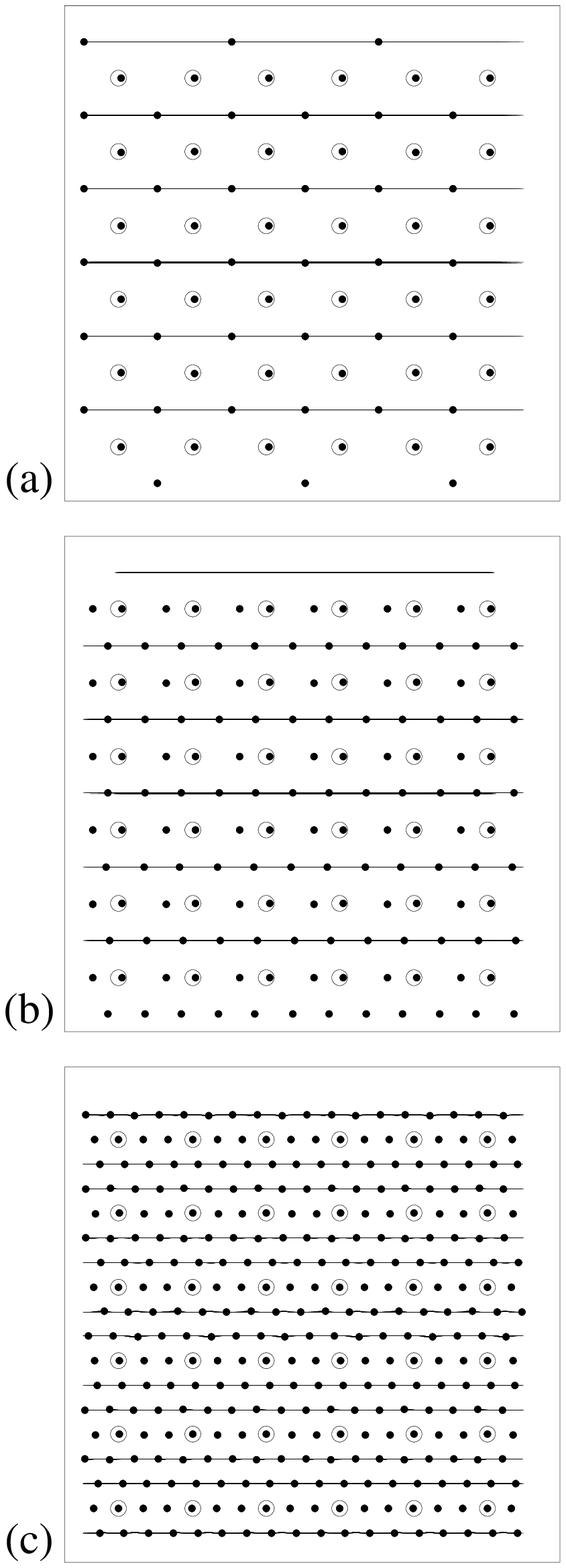}}
\caption{
Vortex trajectories
for the system in Fig.~2 just above depinning for
(a) $B/B_{\phi} = 2$, showing the 1D flow of single rows of 
interstitial vortices. 
(b) $B/B_{\phi} = 4$, showing the 1D flow of 2 rows of 
vortices between pinning sites. Between each pair of pinning sites is
a single immobile interstitial vortex. 
(c) $B/B_{\phi} = 9$, showing the 1D flow where 3 rows of vortices
flow between the pinning sites while a pair of 
immobile interstitial vortices are located between the pinning sites.} 
\label{fig3}
\end{figure}

\begin{figure}
\center{
\epsfxsize=3in
\epsfbox{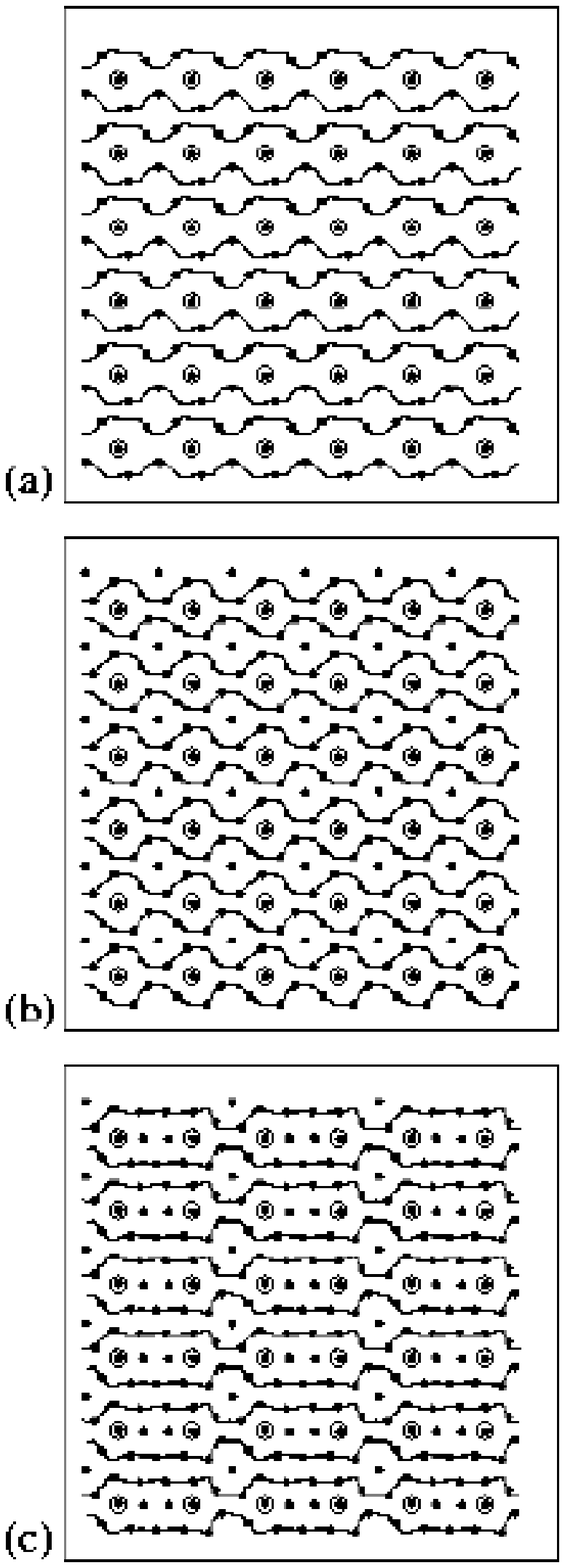}}
\caption{
Vortex trajectories just above depinning for
(a) $B/B_{\phi} = 5$ showing that the flow occurs with two periodic 
winding rows of mobile vortices moving between each pinning row. 
(b)$B/B_{\phi} = 8$ showing a more sinusoidal like flow where again two
rows of vortices move between the pinning rows. In addition, in the center
of each pinning plaquette is an immobile interstitial vortex. 
(c) $B/B_{\phi} = 8.5$ shows a pinched flow. 
An immobile interstitial vortex is located in the center of
every other plaquette and a pair of immobile 
interstitial vortices are located
between every other pair of pinning sites.}
\label{fig4}
\end{figure} 

\begin{figure}
\center{
\epsfxsize=3.5in
\epsfbox{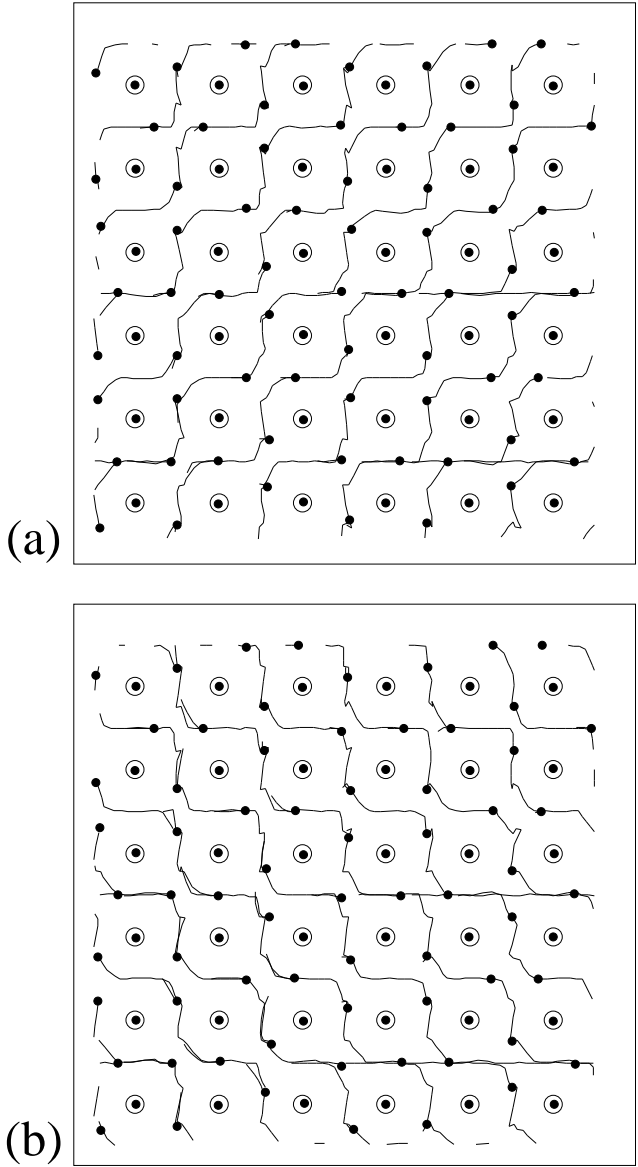}}
\caption{
(a) Vortex trajectories for $B/B_{\phi} = 3$ 
 showing the periodic winding vortex motion
at $+45^{\circ}$. (b) The vortex trajectories for the same system 
and time interval as in (a) after the vortex motion switches to the
$-45^{\circ}$.} 
\label{fig5}  
\end{figure}

\begin{figure}
\center{
\epsfxsize=3.5in
\epsfbox{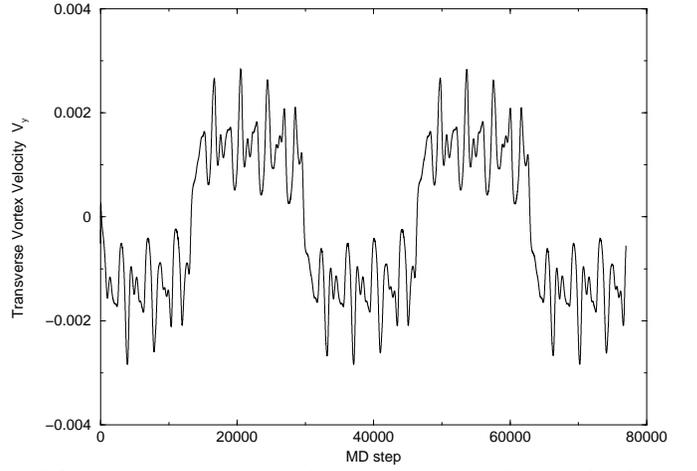}}
\caption{
Transverse, or $y$-direction, vortex velocities for the system in 
Fig.~5 where the driving is in the $x$-direction. 
The vortex motion locks into
the $\pm 45^{\circ}$ direction for 150000 MD steps 
before switching to the
other direction. The additional smaller scale periodic component is due
to the winding nature of the 
vortex flow in the channels as seen in Fig.~5.} 
\label{fig6}
\end{figure}

\begin{figure}
\center{
\epsfxsize=3.5in
\epsfbox{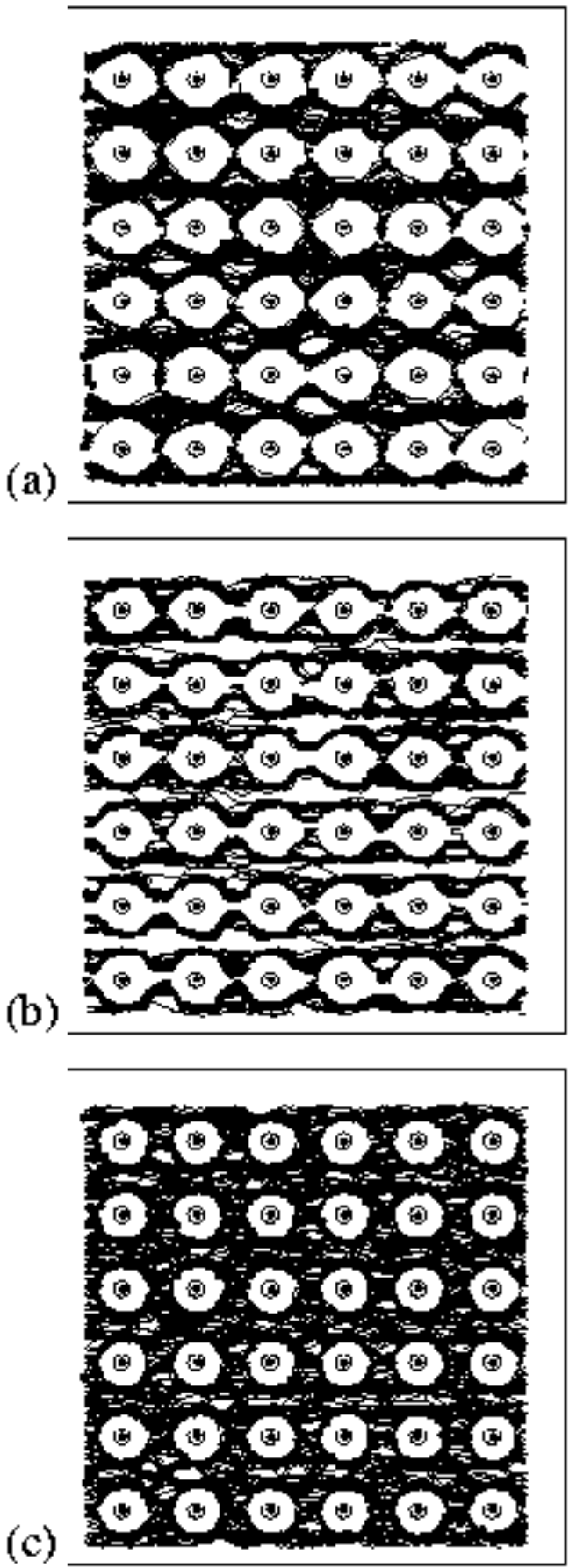}}
\caption{
Vortex trajectories
for (a) $B/B_{\phi} = 4.5$ showing a disordered flow
where the trajectories change over time.
Vortices do not move through the regions near the
occupied pinning sites due to the
vortex-vortex repulsion. (b) $B/B_{\phi} = 4.95$, showing that flow is 
more ordered with the same features of the periodic
winding channels observed 
at $B/B_{\phi} = 5$. If the trajectories are drawn for a longer time 
the plots become increasingly smeared and the plot will become
indistinguishable from that seen in (a). (c) The vortex trajectories for the
same time interval as in (a) and (b) for $B/B_{\phi} = 7$ showing a
disordered flow pattern.}
\label{fig7}
\end{figure} 
 
\begin{figure}
\center{
\epsfxsize=3.5in
\epsfbox{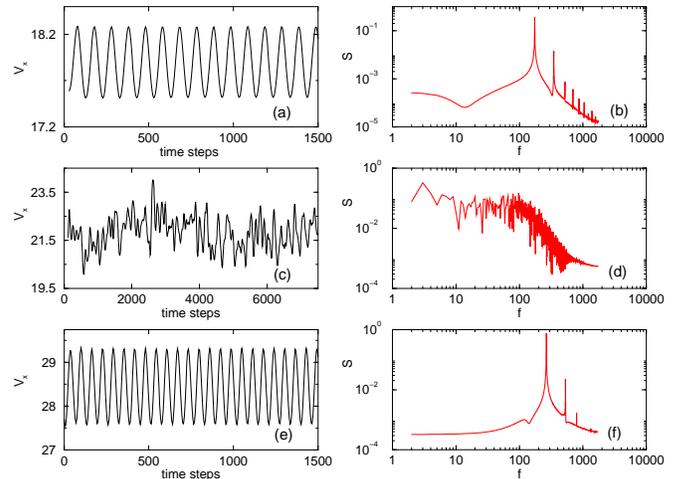}}
\caption{
Time series of the sum of longitudinal vortex velocities
and Fourier transforms for 
the vortex flow. (a,b) $B/B_{\phi} = 4$ where a periodic signal occurs 
due to the 1D flow of vortices (Fig.~3(b)) 
through the periodic potential created by
the pinned vortices. (c,d) $B/B_{\phi} = 4.5$ where a broad spectra occurs 
due to the disordered flow as seen in Fig.~7(a). (e,f) $B/B_{\phi} = 5$
where again a periodic signal occurs as the vortices move in periodic 
winding channels as seen in Fig~4(a). The FT's were taken on data sets of
30000 MD steps.}
\label{fig8}
\end{figure} 

\begin{figure}
\center{
\epsfxsize=3.5in
\epsfbox{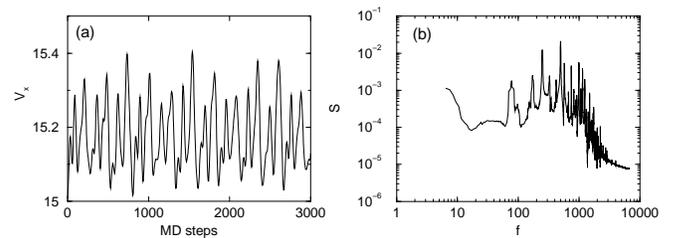}}
\caption{ 
Time series of the sum of the longitudinal vortex
velocities and Fourier transform for the vortex flow 
at $B/B_{\phi} = 3$, where the flow occurs in winding channels.
Here several different periodicities can be distinguished.}
\label{fig9}
\end{figure} 

\begin{figure}
\center{
\epsfxsize=3.5in
\epsfbox{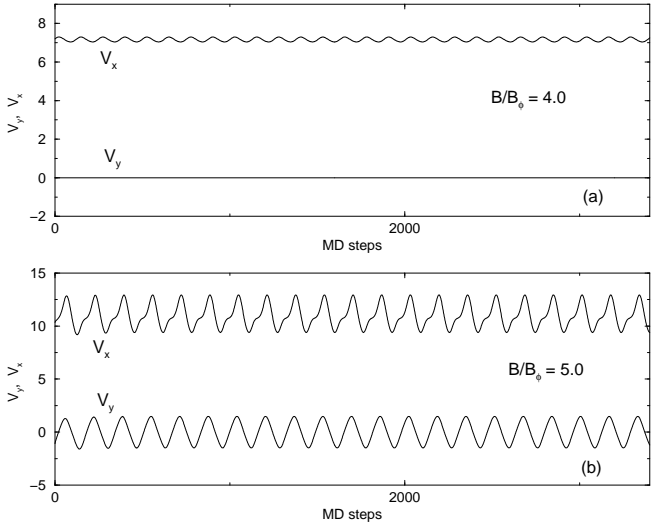}}
\caption{
The time series for the sum of the longitudinal 
velocities $V_{x}$ and
transverse velocities $V_{y}$ for (a) $B/B_{\phi} = 4$ and 
(b) $B/B_{\phi} = 5$ for a fixed drive of $f_{d}/f_{p} = 0.4$. In (a)
where the vortex motion occurs by the 1D flow of interstitial vortices as
seen in Fig.~3(b) the longitudinal velocities show a periodic component while
the transverse velocity is zero. For (b) where the vortex motion was
sinusoidal both the $x$ and $y$ components of the velocities show a
periodic signal.} 
\label{fig10}
\end{figure}

\begin{figure}
\center{
\epsfxsize=3.5in
\epsfbox{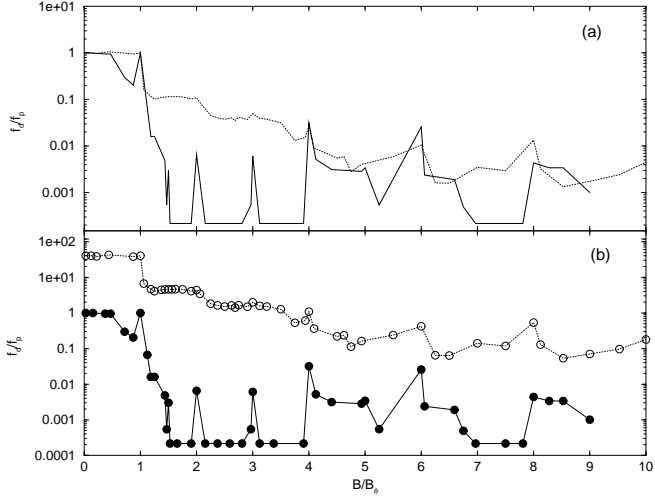}}
\caption{
Critical depinning forces versus $B/B_{\phi}$ 
for driving in the $x$ and $y$ directions for
a system with a rectangular pinning array with $L_{x}/L_{y} = 2$. 
(a) Dashed curve is the depinning line for driving in the $x$-direction
and the solid curve is the depinning line for driving in the $y$-direction. 
For $B/B_{\phi} < 4$ there is a 
clear anisotropy in the depinning force. 
The driving along the $x$-direction gives a higher depinning force 
except at $B/B_{\phi} = 1$,
where the depinning forces both equal $f_{p}$. In (b) the same curve is 
plotted with the $x$-direction driving curve (dashed-line and open circles)
shifted up for clarity. The matching peaks for the $x$-direction 
driving are strongly reduced as compared to the depinning curve for
driving in the $y$-direction. Missing peaks for both curves occur at
$B/B_{\phi} = 5, 7$, and $9$.} 
\label{fig11}
\end{figure}

\begin{figure}
\center{
\epsfxsize=3.5in
\epsfbox{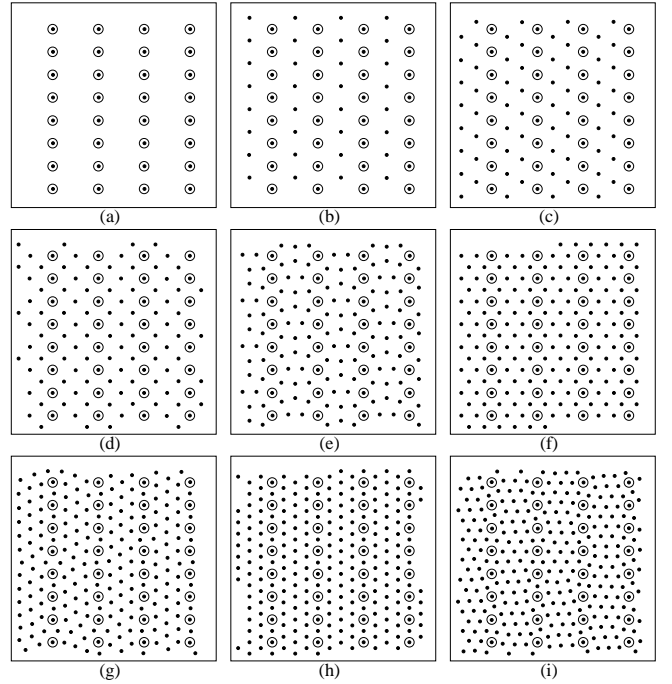}}
\caption{
Vortex positions (black circles) and pinning sites (open circles) 
after annealing from a higher temperature state in a 
rectangular pinning array with $L_{x}/L_{y} = 2$ for 
(a) $B/B_{\phi} = 1$, (b) $2$, (c) $3$, (d) $4$, (e) $5$, 
(f) $6$, (g) $7$, (h) $8$, and (i) $9$.}
\label{fig12}
\end{figure} 

\begin{figure}
\center{
\epsfxsize=3.5in
\epsfbox{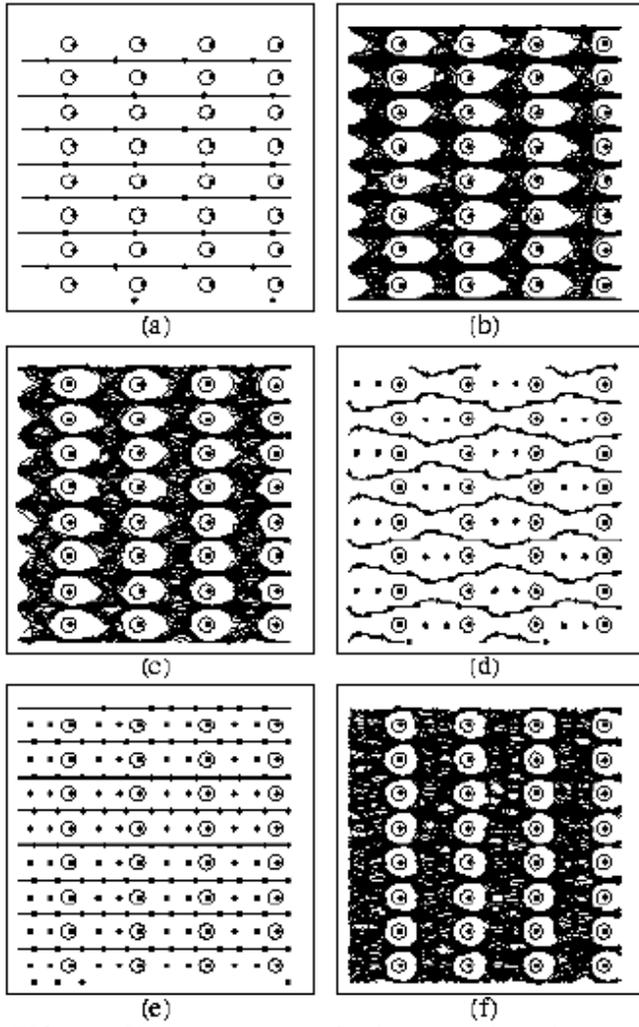}}
\caption{
Vortex trajectories 
for driving in the
$x$-direction for a pinning array with $L_{x}/L_{y} = 2$. 
(a) $B/B_{\phi} = 2$, (b) $3$, 
(c) $4$, (d) $5$, 
(e) $6$, and (f) $9$.} 
\label{fig13}
\end{figure}

\begin{figure}
\center{
\epsfxsize=2.5in
\epsfbox{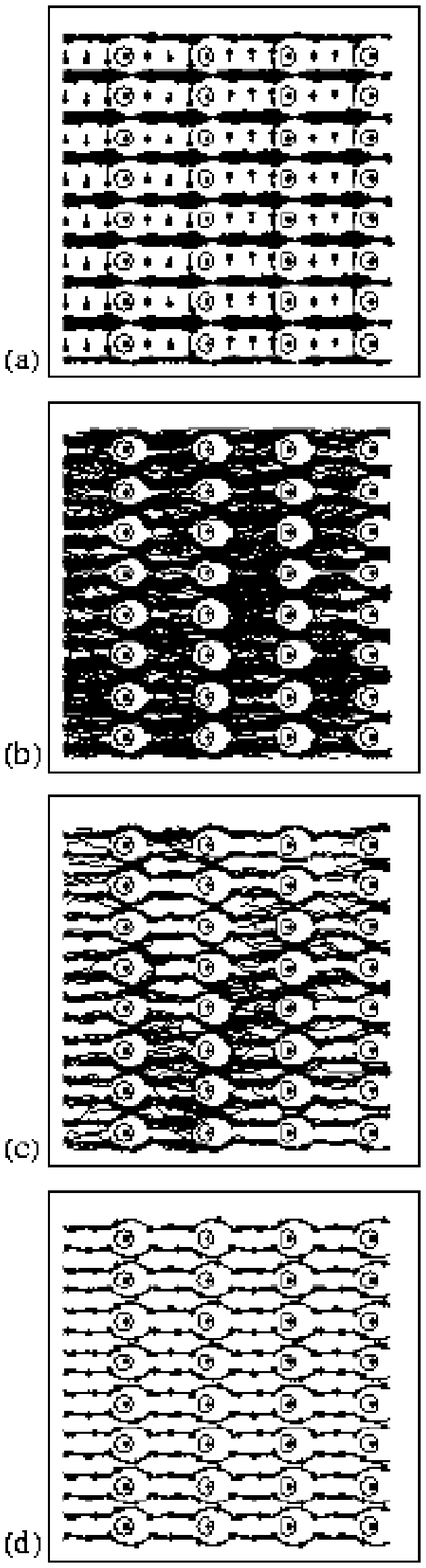}}
\caption{
Vortex trajectories  
for driving in the
$x$-direction in a rectangular pinning array with $L_{x}/L_{y} = 2$
at $B/B_{\phi} = 8$. (a) The flow patterns for $f_{d}/f_{p} = 0.2$, where
the flow is ordered and a portion of the interstitial vortices are immobile. 
(b) $f_{d}/f_{p} = 0.3$ flow becomes disordered.
(c) $f_{d}/f_{p} = 0.35$ flow begins to reorganize.   
(d) $f_{d}/f_{p} = 0.37$ flow is ordered with all the interstitial 
vortices moving.}
\label{fig14}
\end{figure} 

\begin{figure}
\center{
\epsfxsize=3.5in
\epsfbox{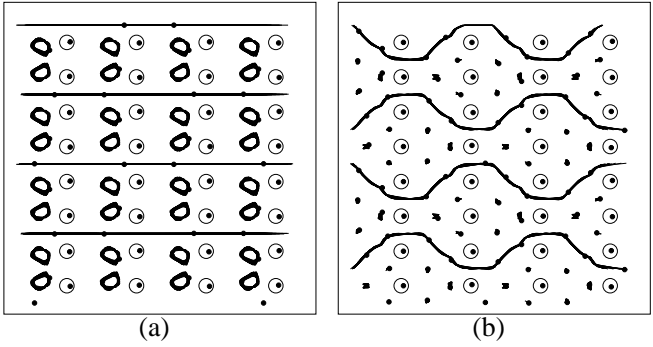}}
\caption{
Vortex trajectories for driving in the $x$-direction
in a rectangular pinning 
array with $L_{x}/L_{y} = 2$ just above depinning. 
(a) $B/B_{\phi} = 2.5$; Vortex motion consists of the 1D flow of interstitial 
vortices
between every other pinning row and the rotational motion of pairs of
interstitial vortices. (b) $B/B_{\phi} = 4.5$; Motion consists
of winding channels that weave between the pinning sites.}
\label{figure15}
\end{figure}
 
\begin{figure}
\center{
\epsfxsize=3.5in
\epsfbox{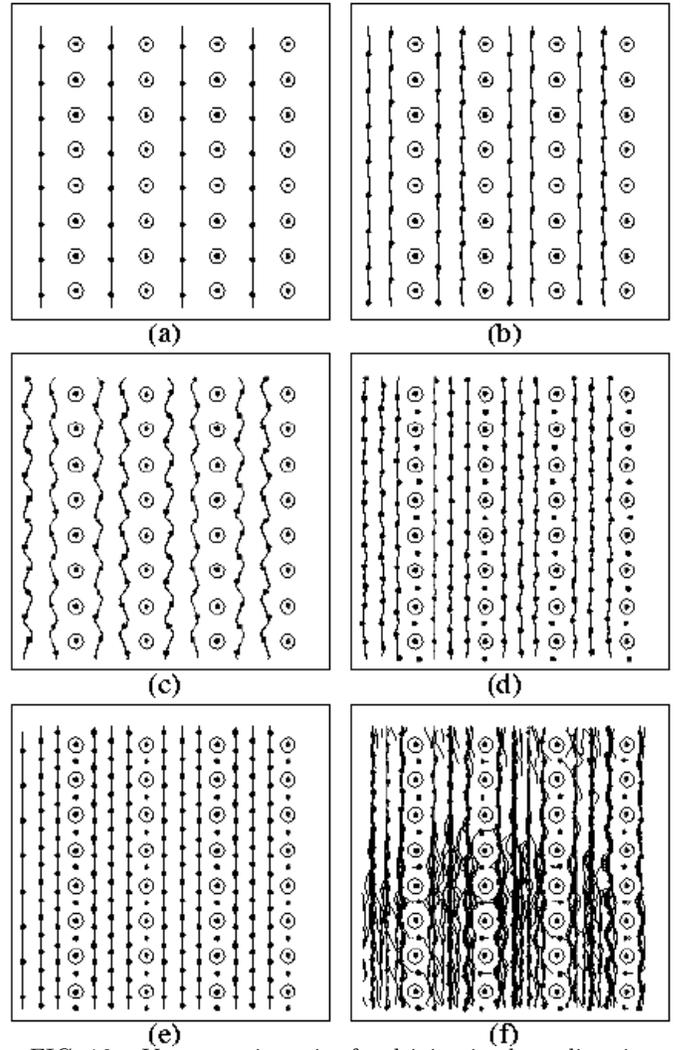}}
\caption{
Vortex trajectories  
for driving in the 
$y$-direction in a rectangular pinning array with $L_{x}/L_{y} = 2$, 
just above depinning. (a) $B/B_{\phi} = 2$, (b) $3$, (c) $4$, 
(d) $7$, (e) $8$ and (f) $9$.} 
\label{figure16}
\end{figure}

\begin{figure}
\center{
\epsfxsize=3.5in
\epsfbox{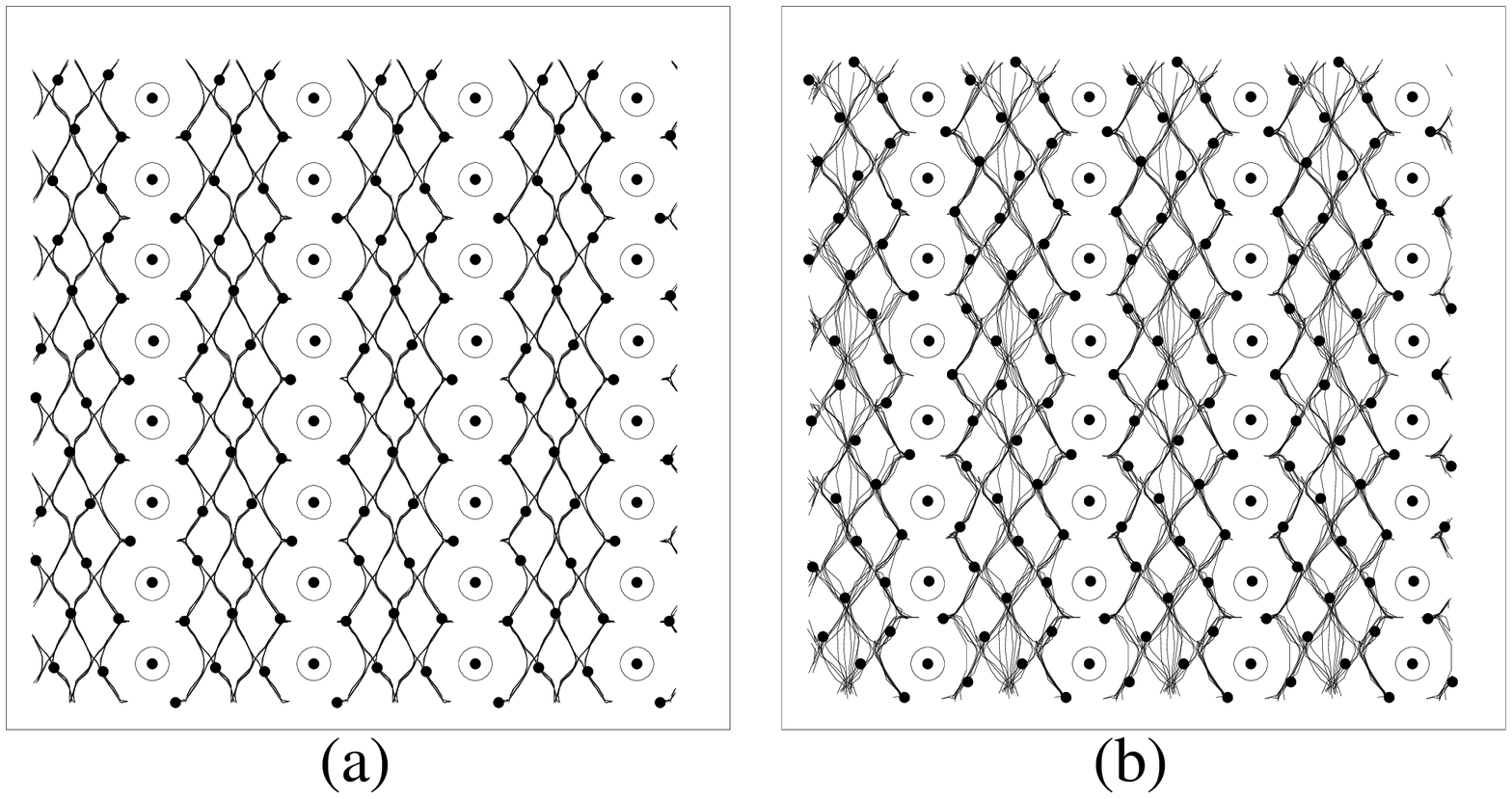}}
\caption{
Vortex trajectories for driving in the $y$-direction
in a rectangular pinning array with $L_{x}/L_{y} = 2$, just above 
depinning. (a) $B/B_{\phi} = 5$ shows an intricate braiding flow pattern.  
(b) $B/B_{\phi} = 6$ shows a similar pattern as in (a) but with some
disorder.}
\label{figure17}
\end{figure} 

\begin{references}

\bibitem{Fiory}
A.T.~Fiory, A.F.~Hebard, and S.~Somekh, Appl.~Phys.~Lett.~{\bf 32}, 73 (1978).

\bibitem{Metlushk}
V.~Metlushko {\it et al.}, Solid State Commun.~{\bf 91}, 331 (1994).

\bibitem{Metlushko}
M.~Baert {\it et al.}, Phys.~Rev.~Lett.~{\bf 74}, 3269 (1995);
V.V.~Moshchalkov {\it et al.}, Phys.~Rev.~B {\bf 54}, 7385 (1996);
V.V.~Moshchalkov {\it et al.}, Phys.~Rev.~B, 3615 (1998);
V.~Metlushko {\it et al.}, Phys.~Rev.~B {\bf 59}, 603 (1999).

\bibitem{Fractional}
M.~Baert {\it et al.}, Europhys.~Lett.~{\bf 29}, 157 (1995).

\bibitem{Wodenweber}
J.Y.~Lin {\it el al.}, Phys.~Rev.~B {\bf 54}, R12714 (1996);
A.~Castellanos {\it et al.}, Appl.~Phys.~Lett.~{\bf 71}, 962 (1997). 

\bibitem{Surdeanu}
R.~Surdeanu, R.J.~Wijngaarden, R.~Wordenweber, and R.~Griessen, to be
published. 

\bibitem{Bezryadin}
A.~Bezryadin, Yu.~B.~Ovchinnikov, and B.~Pannetier, Phys.~Rev.~B
{\bf 53}, 8553 (1996).

\bibitem{Moschalkov}
E.~Rosseel {\it et al.}, Phys.~Rev.~B {\bf 53}, R2983 (1996).

\bibitem{Harada}
K.~Harada {\it et al.}, Science {\bf 271}, 1393 (1996).

\bibitem{Field}
S.~Field {\it et al.}, cond-mat/0003415; A.N.~Grigorenko {\it et al.}, to
be published. 

\bibitem{Bael}
L.~Van Look {\it et al.}, Phys.~Rev.~B {\bf 60}, R6998 (1999).

\bibitem{Welp}
V.~Metlushko {\it et al.}, Phys.~Rev.~B {\bf 60}, R12 585 (1999). 

\bibitem{Schuller}
J.I.~Mart\'{\i}n {\it et al.}, Phys.~Rev.~Lett.~{\bf 79}, 1929 (1997);
Y.~Jaccard {\it et al.}, Phys.~Rev.~B {\bf 58}, 8232 (1998);
A.~Hoffman, P.~Prieto and I.K.~Schuller, Phys.~Rev.~B {\bf 61}, 6958 (2000). 

\bibitem{Jaccard}
D.J.~Morgan and J.B.~Ketterson, Phys.~Rev.~Lett.~{\bf 80}, 3614 (1998). 

\bibitem{Fasano}
Y.~Fasano {\it et al.}, Phys.~Rev.~B {\bf 60}, R15047 (1999).

\bibitem{Terentiev}
A.~Terentiev, D.B.~Watkins, L.E.~De Long, D.J.~Morgan and J.B.~Ketterson, 
Physica C {\bf 324}, 1 (1999); A.~Terentiev {\it et al.}, 
Phys.~Rev.~B {\bf 61}, R9249 (2000). 

\bibitem{DeLong2}
V.V.~Metlushko {\it et al.}, Europhysics Letters {\bf 41}, 333 (1998). 

\bibitem{Hoffman}
J.I.~Mart\'{\i}n {\it et al.}, Phys.~Rev.~Lett.~{\bf 83}, 1022 (1999).

\bibitem{Velez}
J.I.~Mart\'{\i}n {\it et al.}, Phys.~Rev.~B {\bf 62}, 9110 (2000).

\bibitem{Rectangular}
L.~Van Look {\it et al.}, to be published. 

\bibitem{Koles}
S.~Kolesnik {\it et al.}, to be published.

\bibitem{WelpR}
V.V.~Metlushko {\it et al.}, to be published. 

\bibitem{Reichhardt}
C.~Reichhardt, C.J.~Olson, and F.~Nori, Phys.~Rev.~B {\bf 57}, 
7937 (1998).

\bibitem{DrivenShort}
C.~Reichhardt, C.J.~Olson, and F.~Nori, Phys.~Rev.~Lett.~{\bf 78},
2648 (1997).

\bibitem{Shapiro2}
C.~Reichhardt, R.T.~Scalettar, G.T.~Zim\'anyi, and N.~Gr{\o}nbech-Jensen, 
Phys.~Rev.~B R11 914 (2000).

\bibitem{Marconi}
V.I.~Marconi and D.~Dom\'{\i}nguez, Phys.~Rev.~Lett.~{\bf 82}, 4922 (1999);
cond-mat/0008125. 

\bibitem{Janin}
K.D.~Fisher, D.~Stroud, and L.~Janin, Phys.~Rev.~B {\bf 60}, 15371 (1999).

\bibitem{Kes}
A.~Pruyjmboom {\it et al.}, Phys.~Rev.~Lett.~{\bf 60}, 1430 (1988);
M.H.~Theunissen {\it et al.}, {\it ibid}, {\bf 77}, 159 (1996).

\bibitem{Anders}
S.~Anders {\it et al.}, Physica C 332, 35 (2000); and to be published.

\bibitem{Besseling}
R.~Besseling, R.~Niggebrugge, and P.H.~Kes, Phys.~Rev.~Lett.~{\bf 82}, 3144 
(1999).

\bibitem{Higgins}
S.~Bhattacharya and M.J.~Higgins, Phys.~Rev.~Lett.~{\bf 70}, 2617 (1993);
M.J.~Higgins and S.~Bhattacharya, Physica C {\bf 257}, 232 (1996);
M.C.~Hellerqvist {\it et al.}, Phys.~Rev.~Lett.~{\bf 76}, 4022 (1996);
W.~Henderson {\it et al.}, Phys.~Rev.~Lett.~{\bf 77}, 2077 (1996).
 
\bibitem{Marley}
A.C.~Marley, M.J.~Higgins and S.~Bhattacharya, Phys.~Rev.~Lett.~{\bf 74},
3029 (1995).

\bibitem{Yaron}
U.~Yaron {\it et al.}, Nature {\bf 376}, 753 (1995).

\bibitem{Durate}
A.~Duarte {\it et al.}, Phys.~Rev.~B {\bf 53}, 11336 (1996);
F.~Pardo {\it et al.}, Phys.~Rev.~Lett.~{\bf 78}, 2644 (1997). 
 
\bibitem{Pardo}
F.~Pardo {\it et al.}, Nature {\bf 396}, 348 (1998).

\bibitem{Troya}
A.M.~Troyanovski, J.~Aarts, and P.H.~Kes, Nature {\bf 399}, 665 (1999).

\bibitem{Koshelev}
A.E.~Koshelev and V.M.~Vinokur, Phys.~Rev.~Lett.~{\bf 73}, 3580 (1994).

\bibitem{Moon}
K.~Moon, R.T.~Scalettar, and G.T.~Zim\'anyi, Phys.~Rev.~Lett.~{\bf 77},
2278 (1996); S.~Ryu {\it et al.}, Phys.~Rev.~Lett.~{\bf 77}, 5114 (1996).

\bibitem{Dominguez}
H.J.~Jensen, A.~Brass and A.J.~Berlinsky, Phys.~Rev.~Lett.~{\bf 60}, 1676
(1988); A.Ch-Shi and A.J.~Berlinsky, Phys.~Rev.~Lett.~{\bf 67}, 1926 (1991);
N.~Gr{\o}nbech-Jensen, A.R.~Bishop, and D.~Dom\'{\i}nguez, Phys.~Rev.~Lett.~{\bf 76},
2985 (1996); M.C.~Faleski, M.C.~Marchetti, and A.A.~Middleton, 
Phys.~Rev.~B {\bf 54} 12427 (1996); S.~Spencer and H.J.~Jensen, 
Phys.~Rev.~B {\bf 55}, 8473 (1997); D.~Dom\'{\i}nguez, Phys.~Rev.~Lett.~{\bf 82},
181 (1999); A.B.~Kolton, D.~Dom\'{\i}nguez, N.~Gr{\o}nbech-Jensen, 
Phys.~Rev.~Lett.~{\bf 83}, 3061 (1999).

\bibitem{Olson}
C.J.~Olson, C.~Reichhardt, and F.~Nori, Phys.~Rev.~Lett.~{\bf 81}, 3757
(1998).

\bibitem{Giamarchi}
T.~Giamarchi and P.~Le Doussal, Phys.~Rev.~Lett.~{\bf 76}, 3580 
(1996); {\bf 78}, 752 (1997); 
P.~Le Doussal and T.~Giamarchi, Phys.~Rev.~B {\bf 57}, 11 356 (1998);
L.~Balents, M.C.~Marchetti, and L.~Radzihovsky, 
Phys.~Rev.~Lett.~{\bf 78}, 751 (1997); 
Phys.~Rev.~B {\bf 57},
7705 (1998); S.~Scheidl and V.M.~Vinokur, Phys.~Rev.~E {\bf 57}, 2574 (1998). 

\bibitem{Jensen}
N.~Gr{\o}nbech-Jensen, Int.~J.~Mod.~Phys.~C {\bf 7}, 873 (1996);
N.~Gr{\o}nbech-Jensen, Comp.~Phys.~Comm.~{\bf 119}, 115 (1999).

\end{references}
\end{document}